\documentclass{article}
\usepackage{graphicx,caption}
\usepackage[utf8]{inputenc}
\usepackage[a4paper]{geometry}
\usepackage{amsmath,amssymb,amsthm}
\usepackage{bm}
\usepackage{braket}
\usepackage{color}
\usepackage{enumerate}
\usepackage{empheq}
\usepackage{booktabs}
\usepackage{multirow}
\usepackage{quantikz}
\usepackage{comment}

\newtheorem{theorem}{Theorem}

\newtheorem{definition}{Definition}

\begin{document}

\title{Efficient state preparation for\\multivariate Monte Carlo simulation}
\author{Hitomi Mori \thanks{Graduate School of Engineering Science, Osaka University, 1-3 Machikaneyama, Toyonaka, Osaka 560-8531, Japan. \texttt{hmori.academic@gmail.com}} \and Kosuke Mitarai \thanks{Graduate School of Engineering Science, Osaka University, 1-3 Machikaneyama, Toyonaka, Osaka 560-8531, Japan.} \and Keisuke Fujii \thanks{Graduate School of Engineering Science, Osaka University, 1-3 Machikaneyama, Toyonaka, Osaka 560-8531, Japan./ Center for Quantum Information and Quantum Biology, Osaka University, 1-3 Machikaneyama, Toyonaka, Osaka 560-8531, Japan./ Center for Quantum Computing, RIKEN, Hirosawa 2-1, Wako, Saitama 351-0198, Japan./ Fujitsu Quantum Computing Joint Research Division at QIQB,
Osaka University, 1-2 Machikaneyama, Toyonaka 560-0043, Japan}}
\maketitle
\abstract{
Quantum state preparation is a task to prepare a state with a specific function encoded in the amplitude, which is an essential subroutine in many quantum algorithms. In this paper, we focus on multivariate state preparation, as it is an important extension for many application areas. Specifically in finance, multivariate state preparation is required for multivariate Monte Carlo simulation, which is used for important numerical tasks such as risk aggregation and multi-asset derivative pricing. Using existing methods, multivariate quantum state preparation requires the number of gates exponential in the number of variables $D$. For this task, we propose a quantum algorithm that only
requires the number of gates linear in $D$. Our algorithm utilizes multivariable quantum signal processing (M-QSP), a technique to perform the multivariate polynomial transformation of matrix elements. Using easily prepared block-encodings corresponding to each variable, we apply M-QSP to construct the target function. In this way, our algorithm prepares the target state efficiently for functions achievable with M-QSP.
}

\section{Introduction}\label{sec1}
Quantum computer is expected to outperform classical computer in many tasks, including factoring \cite{shor_1994}, Hamiltonian simulation \cite{lloyd_1996}, and matrix inversion \cite{harrow_2009}. Specifically, it is known that Monte Carlo simulation, which is a ubiquitous computational method that appears in diverse fields, such as mathematical finance \cite{Glasserman_2003}, statistical physics \cite{Krauth_2006}, and engineering \cite{Jacoboni_1989}, can obtain quadratic speed-up over the classical counterpart \cite{Montanaro_2015}. However, this speed-up is dependent on a subroutine called quantum state preparation. Quantum state preparation is a task to prepare a superposition with a specific function encoded in the amplitude, and, in general, the upper bound for the gate count is $O(2^n)$ for the $n$-qubit system \cite{mottonen2004}.
If quantum state preparation requires exponentially large cost, it can diminish the quantum speed-up \cite{Herbert_2021}.
Therefore, quantum state preparation is a crucial component in quantum algorithms to achieve a quantum speed-up.
This concern especially becomes prominent when we increase the number of variables for multivariate Monte Carlo simulation.

In this paper, we focus on multivariate state preparation, which encodes a multivariate function in the amplitude. Not only for multivariate Monte Carlo simulation, it has important applications in many areas, including quantum chemistry \cite{Chan_2023} and quantum simulation \cite{Jordan_2012}. Specifically in finance, multivariate Monte Carlo simulation is used for important numerical tasks such as risk aggregation and multi-asset derivative pricing. Risk aggregation is basically a risk analysis on multiple risk variables, where the goal is to quantify the combined risk \cite{Mori_2024}. Multi-asset derivative pricing is a task to price the derivative of multiple underlying assets \cite{Kubo_2022}.

To define the task of the state preparation, we start with the single variable case. The goal of single variable quantum state preparation for Monte Carlo simulation is to prepare the state
\begin{align}
    \ket{\psi}=\frac{1}{\mathcal{N}_p}\sum_{i=0}^{N-1}\sqrt{p(x^{(i)})}\ket{i},\label{stateprep}
\end{align}
where $p$ is the probability density function (PDF) $p:\mathbb{R}\rightarrow\mathbb{R}_{\ge0}$, $x^{(i)}$ is the $i$-th point of $n$-qubit discretization of $x$, $N=2^n$, and $\mathcal{N}_p=\sqrt{\sum_i p(x^{(i)})}$. For this task, various algorithms have been proposed to date, where the multivariable extension of existing methods is summarized in Table \ref{tab1}. 

The Grover-Rudolph method \cite{grover2002creating} encodes the probability distribution by an iterative process, where a qubit is added in each iterate and increases the granularity of the distribution. In each step, an access to the oracle $O_\theta:\ket{i}\ket{0}\mapsto\ket{i}\ket{\theta(x^{(i)})}$ is required for some function $\theta$ that includes the integral of $f$. Therefore, this method is only valid for efficiently integrable distribution functions. There are other methods that avoids the integral. Adiabatic state preparation \cite{rattew2022preparing} starts from the uniform superposition state and adiabatically evolves toward the target distribution state. To implement the adiabatic time evolution, the access to the oracle $O_
\theta$ is assumed. In black-box method \cite{PhysRevLett.85.1334,Sanders_2019,Bausch2022fastblackboxquantum}, the target function is encoded in qubits first using the oracle $O_\theta$ and then exported to the amplitude. When extended to $D$-variable case, these methods rely on $O_\theta$ become much more costly in general due to the cost for the implementation of the oracle. If we use degree $\tilde{d}$ piecewise polynomial approximation following Ref. \cite{häner2018optimizing}, $O(g^2\tilde{d}^D)$ gates and $O(g\tilde{d}^D)$ ancilla qubits are needed to encode $g$-qubit discretization of the function.

Unlike the methods introduced so far, Fourier series loader (FSL) \cite{Moosa_2023} is a state preparation method that does not depend on $O_\theta$. In FSL, the coefficients of the Fourier series approximation of the target function are loaded first, and the Fourier series is obtained through quantum Fourier transform. However, the gate count for FSL also increases rapidly due to the coefficient loading process, where $O(d^D)$ gates are required for degree-$d$ Fourier series. Similarly, another method based on linear combination of unitary (LCU) is recently proposed as a multivariate function state preparation method that avoids arithmetics \cite{rosenkranz2024quantum}. This method implements block-encodings of each term of Fourier or Chebyshev series and combines them using LCU. Because this method also requires to load coefficients of Fourier/ Chebyshev series, $O(d^D)$ gates are necessary for degree-$d$ series.

In this work, we propose a multivariable extension of quantum signal processing (QSP)-based state preparation, which does not require $O_\theta$ and only requires the number of gates linear in $D$. Our method utilizes multivariable quantum signal processing (M-QSP), a technique to perform the multivariate polynomial transform of matrix elements. M-QSP was originally proposed by Ref. \cite{Rossi2022multivariable} and revised by Ref. \cite{németh2023variants,mori2023comment}. For those functions achievable with M-QSP, this method provides an efficient algorithm for multivariable state preparation. Currently known necessary conditions are summarized in Theorem \ref{mqsp}. Even though M-QSP remains under investigation, a program is available to determine if a given set of polynomials can be constructed with M-QSP \cite{ito2024polytime}.
 
In our method, using easily prepared block-encodings representing each variable, we can implement a unitary operator that performs M-QSP in the subspace to construct the polynomial approximation of the target function. Applying this unitary to an initial state, we obtain the superposition of the target state with the amplitude proportional to the filling ratio $\mathcal{F}$, the quantity determined by the integral of the function relative to the maximum value. Lastly, we carry out amplitude amplification $O(1/\mathcal{F})$ times to obtain the target state. Because the block-encodings can be implemented with $O(n)$ gates and the unitary operator requires $O(dD)$ access to the block-encodings, this state preparation method requires $O(ndD/\mathcal{F})$ gates in total, which is linear in $D$.

The rest of this paper is organized as follows. In Section \ref{sec2}, we introduce preliminary knowledge on QSP, M-QSP, and quantum Monte Carlo simulation. The QSP-based state preparation for multivariable functions is explained in Section \ref{sec3}, where the single variable case originally proposed in ref. \cite{mcardle2022quantum} is covered first. As the application of multivariable state preparation, we discuss the quantum algorithm for multivariate Monte Carlo simulation in Section \ref{sec4}. Lastly, in Section \ref{sec5}, we give a conclusion.

\begin{table}[ht]
\begin{tabular*}{\textwidth}{@{\extracolsep\fill}lccc}
\toprule%
Methods  & Gate count & Ancilla qubits  & Function type \\
\midrule
M-QSP-based   & \multirow{2}{*}{$O\left(\frac{nd D}{\mathcal{F}}\right)$} & \multirow{2}{*}{$1$} & Functions achievable\\
(This work)&   &   & with M-QSP (definite parity)\\

\multirow{2}{*}{Grover-Rudolph \cite{grover2002creating}}   & \multirow{2}{*}{$O(nT_{\mathrm{oracle}})$}  &  \multirow{2}{*}{$O(t_{\mathrm{oracle}})$}  & \begin{tabular}{c}Efficiently integrable  \end{tabular}\\
  &   &   & distribution functions \\
  
\multirow{2}{*}{Adiabatic \cite{rattew2022preparing}}   &  \multirow{2}{*}{$O\left(\frac{T_{\mathrm{oracle}}}{\mathcal{F}^4}\right)$} & \multirow{2}{*}{$O(t_{\mathrm{oracle}})$}  & \multirow{2}{*}{Arbitrary functions} \\
&   &   &  \\

\multirow{2}{*}{Black-box \cite{PhysRevLett.85.1334,Sanders_2019,Bausch2022fastblackboxquantum}}   & \multirow{2}{*}{$O\left(\frac{T_{\mathrm{oracle}}}{\mathcal{F}}\right)$}  & \multirow{2}{*}{$O(t_{\mathrm{oracle}})$}  & \multirow{2}{*}{Arbitrary functions} \\
&   &  &  \\

\multirow{2}{*}{FSL \cite{Moosa_2023}}   &  \multirow{2}{*}{$O(d^D+Dn^2)$} & \multirow{2}{*}{0}   & \multirow{2}{*}{Arbitrary functions} \\
&   &   &  \\

\multirow{2}{*}{LCU-based \cite{rosenkranz2024quantum}}   &  \multirow{2}{*}{$O(d^D+Dn\log d)$} & \multirow{2}{*}{$O(D\log d)$}   & \multirow{2}{*}{Arbitrary functions} \\
&   &   &  \\
\bottomrule
\end{tabular*}
\caption{Comparison of multivariable state preparation methods. The number of gates and ancilla qubits required for $O_\theta$ is represented as $T_{\mathrm{oracle}}$ and $t_{\mathrm{oracle}}$, respectively.}\label{tab1}
\end{table}

\section{Preliminaries}\label{sec2}
In this section, we introduce preliminary knowledge used in the following sections.
\subsection{Notations}\label{sec2-1}
When we express a variable $x\in[0,1]$ with $n$ qubits, for $N=2^n$ and $j\in\{0,1,\cdots,N-1\}$, we represent the variable correspond to the computational basis $\ket{j}$ as $x^{(j)}=\frac{j}{N}$. 
We define the block-encoding of an operator $A$ as follows.\\

\begin{definition}[Block-encoding \cite{Gily_n_2019}] 
Suppose that $A$ is an $s$-qubit operator, $\alpha, \varepsilon \in \mathbb{R}_{+}$and $a \in \mathbb{N}$, then we say that the $(s+a)$-qubit unitary $U$ is an $(\alpha, a, \varepsilon)$-block-encoding of $A$, if
\begin{align}
    \left\|A-\alpha\left(\left\langle\left. 0\right|^{\otimes a} \otimes I\right) U\left(|0\rangle^{\otimes a} \otimes I\right) \| \leq \varepsilon\right.\right..
\end{align}
\end{definition}

\subsection{Quantum signal processing}\label{sec2-2}
QSP is a technique to perform the polynomial transform of matrix elements and is capable of describing different algorithms under its framework. The implementable polynomials are fully characterized by the following necessary and sufficient conditions.\\

\begin{theorem}[Necessary and sufficient conditions for QSP \cite{Gily_n_2019}, modified]
Let $d \in \mathbb{N}$. There exists $\Phi=\left\{\phi_0, \cdots, \phi_d\right\} \in \mathbb{R}^{d+1}$ such that for all $x \in [-\pi,\pi]$ :
\begin{align}
    e^{i \phi_0 \sigma_z} \prod_{k=1}^d \left[
    \begin{pmatrix}
        \cos x & i\sin x \\
        i\sin x& \cos x
    \end{pmatrix}
    e^{i \phi_k \sigma_z}\right]=
    \begin{pmatrix}
        P(x) & Q(x) \\
        -Q^*(x) & P^*(x)
    \end{pmatrix}\label{Uphi}
\end{align}
if and only if
\begin{enumerate}[(i)]
\item $P(x)=\sum_{j=0}^da_j\cos(jx)$ and $Q(x)=\sum_{j=1}^d\tilde{a}_j\sin(jx)$ for $a,\tilde{a}\in\mathbb{C}$
\item $P$ has parity $d(\bmod 2)$ and $Q$ has parity $d(\bmod 2)$ under $x \mapsto x+\pi$
\item For all $x \in [-\pi,\pi]$, the relation $|P|^2+|Q|^2=1$ holds.\\\par
\end{enumerate}\label{qsp}
\end{theorem}

Similarly, M-QSP is a technique to perform the multivariate polynomial transform of matrix elements. However, only necessary conditions are known about M-QSP.\\

\begin{theorem}[Necessary conditions for M-QSP \cite{Rossi2022multivariable,mori2023comment}]
Let $d \in \mathbb{N}$. There exist $\Phi=\left\{\phi_0, \cdots, \phi_d\right\} \in \mathbb{R}^{d+1}$ and $s=\{s_1,\cdots,s_d\}\in\{0,1\}^d$ such that for all $x_1,x_2\in [-\pi,\pi]$ :
\begin{align}
    e^{i \phi_0 \sigma_z} \prod_{k=1}^d \left[
    \begin{pmatrix}
        \cos x_1 & i\sin x_1 \\
        i\sin x_1& \cos x_1
    \end{pmatrix}^{s_k}
    \begin{pmatrix}
        \cos x_2 & i\sin x_2 \\
        i\sin x_2& \cos x_2
    \end{pmatrix}^{1-s_k} e^{i \phi_k \sigma_z}\right]=\begin{pmatrix}
        P(x_1,x_2) & Q(x_1,x_2) \\
        -Q^*(x_1,x_2) & P^*(x_1,x_2)
    \end{pmatrix}\label{Uphi2}
\end{align}
only if
\begin{enumerate}[(i)]
\item $P(x_1,x_2)=\sum_{j=-d_1}^{d_1}\sum_{k=-d_2}^{d_2}c_{jk}e^{ijx_1}e^{ikx_2}$ and $Q(x_1,x_2)=\sum_{j=-d_1}^{d_1}\sum_{k=-d_2}^{d_2}\tilde{c}_{jk}e^{ijx_1}e^{ikx_2}$ for $d_1=|s|$ the Hamming weight of $s$, $d_2=d-d_1$, and $c,\tilde{c}\in\mathbb{C}$
\item $P$ has even parity and $Q$ has odd parity under $(x_1,x_2) \mapsto(-x_1,-x_2)$
\item $P$ and $Q$ have parity $d_1(\bmod 2)$ under $x_1 \mapsto x_1+\pi$ and parity $d_2(\bmod 2)$ under $x_2 \mapsto x_2+\pi$
\item For all $x_1,x_2 \in [-\pi,\pi]$, the relation $|P|^2+|Q|^2=1$ holds
\item For $d_1,d_2\ge1$, $P_{d_1}(x_2)=e^{2i\varphi}Q_{d_1}(x_2)$ and/or $P_{d_2}(x_1)=e^{2i\varphi'}Q_{d_2}(x_1)$, where $\varphi,\varphi'\in\mathbb{R}$ and $P_{d_1}(x_2)$ denotes the single variable Laurent polynomial coefficient of the highest $x_1^{d_1}$ term of $P$, and the same goes for others.\\\par
\end{enumerate}\label{mqsp}
We say a function $f$ can be constructed with M-QSP if there exists $(\Phi,s)$ such that the function $P$ or $Q$ on the right-hand side of Eq. \eqref{Uphi2} becomes $f$.
\end{theorem}

\subsection{Exact amplitude amplification}\label{sec2-3}
In this section, we introduce a technique called exact amplitude amplification. We use this in the state preparation algorithm when we extract a target state from a superposition state.\\

\begin{theorem}[Exact amplitude amplification \cite{mcardle2022quantum}]
Let $U$ be an $n$-qubit unitary, $\Pi$ an $n$-qubit projector, $\ket{\psi}$ an $n$-qubit (normalized) quantum state, and $a\ge0$ such that
\begin{align}
    \Pi U\ket{\bar{0}}=a\ket{\psi},
\end{align}
where $\ket{\bar{0}}$ denotes some $n$-qubit initial state. Let $k:=$ $\left\lceil\frac{\pi}{4 \arcsin (a)}-\frac{1}{2}\right\rceil$, and let $\theta:=\frac{\pi}{4 k+2}$. Suppose that we know the value of $a$ and that $R$ is a single-qubit unitary such that $\langle 0|R| 0\rangle=\frac{\sin (\theta)}{a}$. Let us define $U^{\prime}:=R \otimes U$ and
\begin{align}
    W^{\prime}:=U^{\prime}(2|0\rangle\langle 0|\otimes| \overline{0}\rangle\langle\overline{0}|-I) U^{\dagger}(I-2|0\rangle\langle 0| \otimes \Pi),
\end{align}
then
\begin{align}
    (\ket{0}\bra{0}\otimes\Pi)\left(W^{\prime}\right)^k U^{\prime}|0\rangle|\overline{0}\rangle=|0\rangle|\psi\rangle .
\end{align}
\label{EAA}
\end{theorem}

\subsection{Quantum Monte Carlo simulation}\label{sec2-4}
Monte Carlo simulation is a widely used computational method to perform efficient numerical calculations using random numbers. For random variable $X$ following a PDF $p:\mathbb{R}\rightarrow\mathbb{R}_{\ge0}$, the goal of Monte Carlo simulation is to calculate
\begin{align}
    \mathbb{E}[\theta(X)]=\frac{1}{\mathcal{N}_{{p}}^2}\sum_{i}p(x^{(i)}){\theta}(x^{(i)})\label{expectation}
\end{align}
where $\mathcal{N}_{p}=\sqrt{\sum_{i}p}$ and a function $\theta:\mathbb{R}\to[0,1]$ is determined according to the statistic of interest.

Suppose we have access to a state preparation unitary
\begin{align}
    \mathcal{A}_{{p}}\ket{0}=\frac{1}{\mathcal{N}_{p}}\sum_{i}\sqrt{p(x^{(i)})}\ket{i}
\end{align}
and a controlled rotation
\begin{align}W_\theta:\ket{0}\ket{i}\rightarrow\left(\sqrt{\theta(x^{(i)})}\ket{0}+\sqrt{1-\theta(x^{(i)})}\ket{1}\right)\ket{i},\label{W}
\end{align}
we can estimate $\mathbb{E}[\theta]$ with the following theorem:\\

\begin{theorem}[Quantum Monte Carlo simulation \cite{Montanaro_2015}]
    Let $\delta,\epsilon\in(0,1)$.
    Assuming we have access to the state preparation oracle $\mathcal{A}_{p}$ for a joint pdf $p:\mathbb{R}\rightarrow\mathbb{R}_{\ge0}$ and the controlled rotation oracle $W_{\theta}$ for a function $\theta:\mathbb{R}\rightarrow[0,1]$, there exists a quantum algorithm that, with probability at least $1-\delta$, outputs an $\epsilon$-approximation of ${\mathbb{E}}[\theta(X)]$, querying $\mathcal{A}_{p}$ and $W_\theta$ $O\left(\frac{1}{\epsilon}\log\left(\frac{1}{\delta}\right)\right)$ times each.\\\par
    \label{th:QMCI}
\end{theorem}

There are a number of methods proposed to construct the state preparation oracle $\mathcal{A}_p$, as discussed in Introduction \cite{grover2002creating,rattew2022preparing,PhysRevLett.85.1334,Sanders_2019,Bausch2022fastblackboxquantum,Moosa_2023,rosenkranz2024quantum}. However, these methods require the number of gates exponential in the number of variables $D$, and therefore they are not scalable with respect to $D$. Similarly, for the construction of the controlled rotation $W_\theta$, arithmetic operation is usually assumed, but this approach is not scalable. For single variable case, QSP-based construction of $W_\theta$ was recently proposed by Ref. \cite{stamatopoulos2024quantum}.

\section{M-QSP-based quantum state preparation}\label{sec3}
In this section, we develop the quantum algorithm for state preparation using M-QSP, extending the procedure in Ref. \cite{mcardle2022quantum} (see Appendix \ref{app1}) to multivariable case. For the simplicity of the discussion, we will first focus on two normalized variables $x_1,x_2\in[0,1]$, but the generalization to higher dimension is straightforward and will be explained later.

Our goal for two variable case is to prepare the superposition of 
\begin{align}
    \ket{\psi_f}=\frac{1}{\mathcal{N}_f}\sum_{i=0}^{N_1-1}\sum_{j=0}^{N_2-1}f(x_1^{(i)},x_2^{(j)})\ket{i}\ket{j},
\end{align}
where $\mathcal{N}_f=\sqrt{\sum_{i,j}(f(x_1^{(i)},x_2^{(j)}))^2}$. Here, we assume the number of bits to represent $x_1, x_2$ as $n_1,n_2$ and $N_1=2^{n_1},N_2=2^{n_2}$.

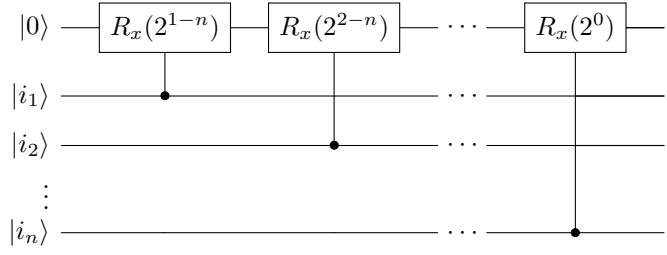
\begin{figure}[t]
\centering
\captionsetup{width=0.7\textwidth}
\begin{quantikz}[thin lines]
    \lstick{$\ket{0}$} & \gate{R_x(2^{1-n})} & \gate{R_x(2^{2-n})} & \ \cdots\ & \gate{R_x(2^{0})} & \qw \\
    \lstick{$\ket{i_1}$} & \ctrl{-1} & & \ \cdots\ & & \qw \\
    \lstick{$\ket{i_2}$} & & \ctrl{-2} & \ \cdots\ & & \qw \\
    \lstick{$\vdots$} & \setwiretype{n} & & & & \\
    \lstick{$\ket{i_n}$} & & &\ \cdots\ &\ctrl{-4} &
\end{quantikz}
\caption{Implementation of $U_1$. Here $\ket{i}$ is represented as $\ket{i_n\cdots i_1}$.}\label{fig_BE}
\end{figure}

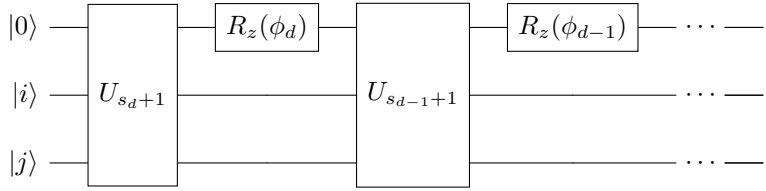
\begin{figure}[t]
\centering
\captionsetup{width=0.7\textwidth}
\begin{quantikz}[thin lines]
    \lstick{$\ket{0}$} & \gate[3]{U_{s_d+1}} & \gate{R_z(\phi_d)} & \gate[3]{U_{s_{d-1}+1}} & \gate{R_z(\phi_{d-1})} & \ \cdots\  & \qw \\
    \lstick{$\ket{i}$} & & & & & \ \cdots\ & \qw \\
    \lstick{$\ket{j}$} & & & & & \ \cdots\ & \qw 
\end{quantikz}
\caption{Implementation of the sequence Eq. \eqref{MQET}.}\label{fig4}
\end{figure}

To achieve our goal, we prepare $(1,1,0)$-block-encodings of $A_1=\sum_i\cos(x_1^{(i)})\ket{i}\bra{i}$ and $A_2=\sum_j\cos(x_2^{(j)})\ket{j}\bra{j}$ as
\begin{align}
    &U_1=\sum_{i}\begin{pmatrix}
    \cos\left(x_1^{(i)}\right)&i\sin\left(x_1^{(i)}\right)\\
    i\sin\left(x_1^{(i)}\right)&\cos\left(x_1^{(i)}\right)\\
    \end{pmatrix}\otimes\ket{i}\bra{i}\otimes I_{n_2},\\
    &U_2=\sum_{j} \begin{pmatrix}
    \cos\left(x_2^{(j)}\right)&i\sin\left(x_2^{(j)}\right)\\
    i\sin\left(x_2^{(j)}\right)&\cos\left(x_2^{(j)}\right)\\
    \end{pmatrix}\otimes I_{n_1}\otimes\ket{j}\bra{j}.
\end{align}
These can be implemented with $n_1$ and $n_2$ controlled X-rotations respectively as in Fig. \ref{fig_BE}.
Then for $\Phi=\left\{\phi_0, \phi_1, \cdots, \phi_n\right\} \in \mathbb{R}^{n+1}$ and $s=\{s_1,\cdots,s_n\}\in\{0,1\}^n$, we construct the sequence
\begin{align}
    U_\Phi=(e^{i\phi_0 \sigma_z}\otimes I_{n_1+n_2}) \prod_{k=1}^d U_1^{1-s_k} U_2^{s_k} (e^{i\phi_k \sigma_z}\otimes I_{n_1+n_2}),\label{MQET}
\end{align}
using $O(d_1)$ $U_1$ and $O(d_2)$ $U_2$, where $d_1=|s|$ and $d_2=d-d_1$ as in Fig. \ref{fig4}.
When projected onto the subspace spanned by $\ket{i}\ket{j}$ of the system registers, operations on the ancilla qubit become 
\begin{align}
    e^{i\phi_0\sigma_z} \prod_{k=1}^d \begin{pmatrix}
    \cos\left(x_1^{(i)}\right)&i\sin\left(x_1^{(i)}\right)\\
    i\sin\left(x_1^{(i)}\right)&\cos\left(x_1^{(i)}\right)\\
    \end{pmatrix}^{1-s_k} \begin{pmatrix}
    \cos\left(x_2^{(j)}\right)&i\sin\left(x_2^{(j)}\right)\\
    i\sin\left(x_2^{(j)}\right)&\cos\left(x_2^{(j)}\right)\\
    \end{pmatrix}^{s_k} e^{i\phi_k\sigma_z},
\end{align}
which is M-QSP. 
When a function 
$f/|f|_{\mathrm{{max}}}$ with a normalization factor $|{f}|_{\mathrm{max}}$ defined as the maximum value of $|{f}|$ can be constructed by M-QSP, we can choose $\Phi$ and $s$ such that Eq. \eqref{MQET} becomes
\begin{align}
    U_\Phi=&\sum_{i,j}
    \begin{pmatrix}
    \frac{{f}(x_1^{(i)},x_2^{(j)})}{|{f}|_{\mathrm{max}}}&\cdot\\
    \cdot&\cdot\\
    \end{pmatrix}\otimes\ket{i}\bra{i}\otimes\ket{j}\bra{j}.
\end{align}
Applying $U_\Phi$ to the initial state $\ket{0}\ket{+}^{\otimes n_1+n_2}$, we obtain
\begin{align}
    \ket{\Psi_{{f}}}&=\sum_{i,j}\frac{{f}(x_1^{(i)},x_2^{(j)})}{\sqrt{N_1N_2}|{f}|_{\mathrm{max}}}\ket{0}\ket{i}\ket{j}+\ket{\perp}\\
    &=\mathcal{F}_{{f}}\ket{0}\ket{\psi_{{f}}}+\ket{\perp},
\end{align}
where the filling ratio in this case is $\mathcal{F}_{{f}}=\mathcal{N}_{{f}}/(\sqrt{N_1N_2}|{f}|_{\mathrm{max}})$ and ($\bra{0}\otimes I_{n_1+n_2})\ket{\perp}=0$. Supposing we know the value of $\mathcal{F}_{{f}}$ and applying exact amplitude amplification (Theroem \ref{EAA}), which requires $O(1/\mathcal{F}_{{f}})$ queries to $U_\Phi$, we obtain $\ket{\psi_{{f}}}$. We have now shown the following:\\

\begin{theorem}[Bivariate state preparation]
Given a Fourier series ${f}$ of degree $(d_{1}, d_{2})$ that can be constructed with M-QSP, we can prepare a quantum state $\ket{\psi_{{f}}}$ using $O\left((n_1d_{1}+n_2d_{2})/\mathcal{F}_{{f}}\right)$gates.\\
\label{BiState}
\end{theorem}

Generalization of the discussion above to $D$-variable function leads to the following theorem: we only have to prepare starting block-encodings for each variable and interleave them in the sequence.\\

\begin{theorem}[Multivariate state preparation]
Given a Fourier series ${f}$ of degree $(d_1,\cdots,d_D)$ that can be constructed with M-QSP, we can prepare a quantum state $\ket{\psi_{{f}}}$ using $O\left(nd D/\mathcal{F}_{{f}}\right)$ gates,
where $n=\max\{n_1,\cdots,n_D\}$ and $d=\max\{d_1,\cdots,d_D\}$.
\label{MultiState}
\end{theorem}

\section{Application to multivariate Monte Carlo simulation}\label{sec4}
In this section, we develop an algorithm for multivariate Monte Carlo simulation extending the discussion in Ref. \cite{Montanaro_2015,Woerner_2019,stamatopoulos2024quantum} to multivariable cases. 

\subsection{Multivariate Monte Carlo simulation}\label{sec4-1}
Monte Carlo simulation is a computational method utilizing random samplings. In multivariate Monte Carlo simulation, several random variables are simulated simultaneously considering the correlation among them. This procedure is important when one needs to combine several random variables, which is the case in risk aggregation and multi-asset derivative pricing. The task can be described as the computation of the expected value of a function $\theta$, which corresponds to the kind of statistic as discussed later. Specifically, for random variable $\bm{X}=(X_1,\cdots,X_D)$, the goal is to calculate
\begin{align}
    \mathbb{E}[\theta(\bm{X})]=\frac{1}{\mathcal{N}^2_{p}}\sum_{\bm{i}}p(x_1^{(i_1)},\dots,x_D^{(i_D)}){\theta}(x_1^{(i_1)},\dots,x_D^{(i_D)})\label{expectation}
\end{align}
for a joint PDF $p:\mathbb{R}^D\rightarrow\mathbb{R}_{\ge0}$.

When the joint PDF $p$ and the function $\theta$ can both be constructed with M-QSP, our method in Section \ref{sec3} can be applied to efficiently prepare both $\mathcal{A}_p$ and $W_\theta$ in Theorem \ref{th:QMCI}.
Suppose a joint PDF $p$ that can be constructed with length-$d_{p}$ M-QSP. We can construct the state preparation unitary
\begin{align}
    \mathcal{A}_{{p}}\ket{0}=\frac{1}{\mathcal{N}_{{p}}}\sum_{\bm{i}}\sqrt{{p}(\bm{x}^{(\bm{i})})}\ket{i_1}\dots\ket{i_D}
\end{align}
for normalized variables $x_1,\cdots,x_D\in[0,1]$, where $\mathcal{N}_{{p}}=\sqrt{\sum_{\bm{i}}{p}}$ and $\bm{x}^{(\bm{i})}=(x_1^{(i_1)},\dots,x_D^{(i_D)})$.
Based on theorem \ref{MultiState}, this state can be prepared with $O(nd_{p}D/\mathcal{F}_{{p}})$, where $\mathcal{F}_{{p}}=\mathcal{N}_{{p}}/\sqrt{N}$ for $N=N_1\cdots N_D$.
Suppose also that the square root of a function ${\theta}:[0,1]\to[0,1]$ can be constructed with length-$d_{\theta}$ M-QSP. We can construct
\begin{align}
    W_{{\theta}}:\ket{0}\ket{i_1}\cdots\ket{i_D}\rightarrow\left(\sqrt{{\theta}(\bm{x}^{(\bm{i})})}\ket{0}+\sqrt{1-{\theta}(\bm{x}^{(\bm{i})})}\ket{1}\right)\ket{i_1}\cdots\ket{i_D}\label{W}
\end{align}
in the same manner to $U_\Phi$ in Section \ref{sec3} using $O(nd_{\theta}D)$ gates. 
Using these $\mathcal{A}_p$ and $W_\theta$, we could perform quantum Monte Calro simulation in Theorem \ref{th:QMCI}.
From the above discussion, we obtain the following result:\\

\begin{theorem}[Multivariate quantum Monte Carlo simulation]
Let $\delta,\epsilon\in(0,1)$. Suppose that the square root of a joint PDF $p$ can be constructed with length-$d_{p}$ M-QSP and the square root of a function ${\theta}:[0,1]\to[0,1]$ can be constructed with length-$d_{\theta}$ M-QSP. A state preparation unitary $\mathcal{A}_{{p}}$ can be obtained with $O(nd_{p}D/\mathcal{F}_{{p}})$ gates, and a controlled rotation $W_{{\theta}}$ can be obtained with $O(n{d}_{\theta} D)$ gates. Using these unitaries, there exists a quantum algorithm that, with probability at least $1-\delta$, outputs an $\epsilon$-approximation of $a=\mathbb{E}[\theta(\bm{X})]$, querying $\mathcal{A}_{{p}}$ and $W_{{\theta}}$ $O\left(\frac{1}{\epsilon}\log\left(\frac{1}{\delta}\right)\right)$ times each.
\label{QMCI}
\end{theorem}

\subsection{Example: Risk aggregation}\label{4-2}
Using Theorem \ref{QMCI}, we can calculate any statistics by appropriately setting the function $\theta$. Here we show one of the most important examples, risk measure, where the common metrics are Value at Risk (VaR) and Tail Value at Risk (TVaR)\footnote{Also known as Conditional Value at Risk (CVaR).}. The goal here is to estimate these measures for the sum of variables $S=(X_1+\cdots+X_D)/D$, where $S\in[0,1]$ is normalized so that we could apply QSP. This task is specifically called risk aggregation.

\begin{figure}[t]
\centering
\captionsetup{width=0.7\textwidth}
\begin{quantikz}[thin lines]
    \lstick{$\ket{0}$} & \gate{R_x(2^{1-n})}  & \ \cdots\ & \gate{R_x(2^{0})} & \gate{R_x(2^{1-n})}  & \ \cdots\ & \gate{R_x(2^{0})} & \ \cdots\ \\
    \lstick{$\ket{i_1}$} & \ctrl{-1} & \ \cdots\ & & & \ \cdots\ & & \ \cdots\ \\
    \lstick{$\vdots$} & \setwiretype{n} \\
    \lstick{$\ket{i_n}$} & &\ \cdots\ &\ctrl{-3} & & \ \cdots\ & & \ \cdots\ \\
    \lstick{$\ket{j_1}$} & & \ \cdots\ & & \ctrl{-4} & \ \cdots\ & & \ \cdots\ \\
    \lstick{$\vdots$} & \setwiretype{n} \\
    \lstick{$\ket{j_n}$} & & \ \cdots\ & & &\ \cdots\ &\ctrl{-6} &\ \cdots\
\end{quantikz}
\caption{Implementation of $U$ for $s=x_1^{(i)}+x_2^{(j)}$.}\label{fig_s}
\end{figure}
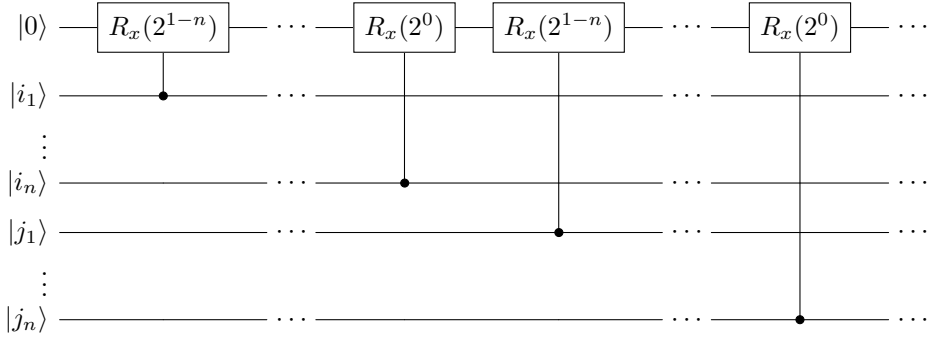

Thus, in risk aggregation, we need a state preparation unitary $\mathcal{A}_p$ for a joint PDF $p(\bm{x})$ and a controlled rotation $W_\theta$ for $\theta(s)$. Specifically, as a joint PDF, simple multivariate distributions such as multivariate normal distribution or more flexible distributions where arbitrary marginal PDFs are substituted into a multivariate function called copula is used. When the distribution can be well approximated by a M-QSP implementable function, $\mathcal{A}_p$ can be prepared with our M-QSP-based method using $O(nd_{p}D/\mathcal{F}_{{p}})$ gates, as discussed in Section \ref{sec4-1}. On the other hand, the construction of $W_\theta$ can be treated as a single variable problem, and thus we only need QSP rather than M-QSP. It is because we can efficiently prepare $(1,1,0)$-block encoding of $A=\sum_{\bm{i}}\cos(s^{(\bm{i})})\ket{\bm{i}}\bra{\bm{i}}$ defined as
\begin{align}
    \sum_{\bm{i}}\begin{pmatrix}
        \cos\left(s^{(\bm{i})}\right)&i\sin\left(s^{(\bm{i})}\right)\\
        i\sin\left(s^{(\bm{i})}\right)&\cos\left(s^{(\bm{i})}\right)
    \end{pmatrix}\otimes\ket{\bm{i}}\bra{\bm{i}}.
\end{align}
It can be constructed by Fig. \ref{fig_s} with $O(nD)$ gates. If we have Chebyshev series approximation of the function $\theta$, we could apply QSP (Theorem \ref{qsp}) to this block encoding, and we could prepare $W_\theta$ with $\theta$ determined by the kind of risk measure. The following strategy will give us an estimate of VaR or TVaR with an additional error $O(\Delta)$ using $O(\frac{nD}{\Delta}\log\frac{1}{\epsilon_\theta})$ gates, where $\Delta$ is a gap parameter to approximate a step function with Chebyshev polynomial.

We start with VaR, which is most common in financial regulations and disclosures. VaR of a random variable $X$ at a confidence level $\alpha\in[0,1]$ is defined as
\begin{align}
    \mathrm{VaR}_X(\alpha):=\inf\{x\in\mathbb{R}:\mathrm{Pr}(X \le x) \ge \alpha\}.
\end{align}
Calculating $\mathrm{VaR}_X(\alpha)$ therefore reduces to the problem of calculating $\mathrm{Pr}(X \le x)$ for a given $x$, because then we can find the smallest $x_\alpha$ such that $\mathrm{Pr}(X\le x_\alpha)\le\alpha$ using bisection search.
Note that, in finance, we are particularly interested in $\mathrm{VaR}_S(\alpha)$.

To estimate $\mathrm{Pr}(S \le l)$ via Theorem \ref{QMCI}, we can set the function $\tilde{\theta}$ as the Chebyshev series approximation of a step function
\begin{align}
    \theta_l(s)=
    \begin{cases}
    1&(s\le l)\\
    0&(s> l)
    \end{cases}.\label{var}
\end{align}
Plugging Eq. \eqref{var} into Eq. \eqref{expectation}, we obtain
\begin{align}
    \frac{1}{\mathcal{N}^2_{{p}}}\sum_{\bm{i}:s\le l} {p}(\bm{x}^{(\bm{i})})=\mathrm{Pr}(S\le l),
\end{align}
as desired. Using the polynomial $P^{\mathrm{rect}}_{\epsilon_\theta,1/\Delta}$ in Ref. \cite{Martyn_2021}, the $\epsilon_\theta$-approximation of Eq. \eqref{var} can be constructed with degree $d_\theta=O(\frac{1}{\Delta}\log\frac{1}{\epsilon_\theta})$\footnote{We could also find a polynomial approximation by solving minimax optimization problem as in Ref. \cite{stamatopoulos2024quantum}}.

TVaR is the conditional expected value of the loss in the tail of the distribution that exceeds the VaR, so it can capture extreme loss events included in the tail. TVaR of $X$ at a confidence level $\alpha$ is expressed as
\begin{align}
    \mathrm{TVaR}_X(\alpha)&:=\mathbb{E}_X[X|X \ge \mathrm{VaR}_X(\alpha)].
\end{align}
For TVaR, using the estimated $\mathrm{VaR}_\alpha(S)$, denoted as $l_\alpha$, we set the function $\tilde{\theta}$ as the Chebyshev series approximation of
\begin{align}
    \theta_{l_\alpha}(s)=
    \begin{cases}
    0&(s\le l_\alpha)\\
    s&(s> l_\alpha).
    \end{cases}
\end{align}
Then the QAE result multiplied by $D/(1-\alpha)$ yields TVaR$_\alpha$. We can calculate the Chebyshev coefficients in the similar manner to VaR, but in this case we can take a parallel shift as
\begin{align}
    \theta_{l_\alpha}(s)=
    \begin{cases}
    0&(s\le l_\alpha)\\
    s-l_\alpha&(s> l_\alpha)
    \end{cases}
\end{align}
to avoid discontinuity \cite{stamatopoulos2024quantum}. As a result, the Chebyshev polynomial approximation in this case requires a much smaller number of terms $d$ compared to VaR calculation, and thus the calculation cost for VaR becomes dominant in TVaR calculation.

Taking the degree of polynomial to approximate $\theta(s)$ in Eq. \eqref{var} as $d_\theta=O(\frac{1}{\Delta}\log\frac{1}{\epsilon_\theta})$, the number of gates required to construct $W_\theta$ for VaR and TVaR can be expressed as $O(\frac{nD}{\Delta}\log\frac{1}{\epsilon_\theta})$ with an additional additive error $O(\Delta)$.
\\

\section{Conclusion}\label{sec5}
We developed an efficient quantum algorithm for multivariate state preparation, which is a crucial subroutine in multivariate Monte Carlo simulation. Preparation of the state with $D$-variable function encoded in the amplitude usually requires the number of qubits exponential in $D$. In our algorithm, we utilize M-QSP, a technique to perform multivariate polynomial transforms of signal operators, and achieve the same goal with the number of qubits linear in $D$. Even though M-QSP remains under investigation, a program is available to determine if a given set of polynomials is achivable with M-QSP \cite{ito2024polytime}.

\section*{Acknowledgement}
This work is supported by MEXT Quantum Leap Flagship Program (MEXT Q-LEAP) Grant No. JPMXS0120319794, JST COI-NEXT Grant No. JPMJPF2014.

\appendix
\section{Quantum state preparation}\label{app1}
In this appendix, we introduce the single variable state preparation method originally proposed by Ref. \cite{mcardle2022quantum}. The goal is to prepare the state
\begin{align}
    \ket{\psi_f}=\frac{1}{\mathcal{N}_f}\sum^{N-1}_{i=0}f(x^{(i)})\ket{i},
\end{align}
where $f:\mathbb{R}\to\mathbb{R}_{\ge0}$ and $\mathcal{N}_f=\sqrt{\sum_{i}(f(x^{(i)}))^2}$. For Monte Carlo simulation, we substitute $f=\sqrt{p}$ for a PDF $p$ as in Eq. \eqref{stateprep}. 
This method utilizes the easily implemented block-encoding and performs QSP on the ancilla qubit to construct the target function, followed by exact amplitude amplification to obtain the target state.

To apply QSP, we first need to classically compute the Chebyshev series approximation 
$\tilde{f}$ of the target function ${f}$ for even and odd part: 
\begin{align}
    &\tilde{f}(x)=\tilde{f}^e(x)+\tilde{f}^o(x),\\
    &\tilde{f}^e(x):=\sum_{l~\mathrm{even}}a_j\cos(lx),\\
    &\tilde{f}^o(x):=\sum_{l~\mathrm{odd}}a_j\cos(lx).
\end{align}
According to Theorem \ref{qsp}, there exist angle parameters $\Phi^{e}$ and $\Phi^{o}$ for each function.

\begin{figure}[t]
\centering
\captionsetup{width=0.7\textwidth}
\begin{quantikz}[thin lines]
    \lstick{$\ket{0}$} & \gate{R_x(2^{1-n})} & \gate{R_x(2^{2-n})} & \ \cdots\ & \gate{R_x(2^{0})} & \qw \\
    \lstick{$\ket{i_1}$} & \ctrl{-1} & & \ \cdots\ & & \qw \\
    \lstick{$\ket{i_2}$} & & \ctrl{-2} & \ \cdots\ & & \qw \\
    \lstick{$\vdots$} & \setwiretype{n} & & & & \\
    \lstick{$\ket{i_n}$} & & &\ \cdots\ &\ctrl{-4} &
\end{quantikz}
\caption{Implementation of $U$.}\label{fig1}
\end{figure}
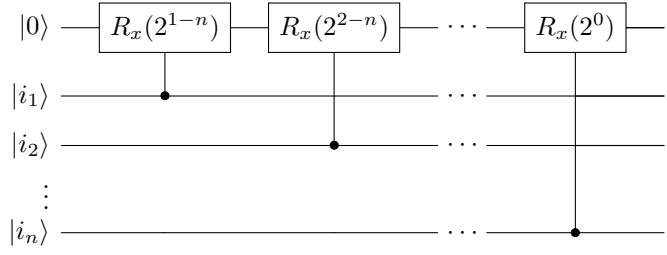

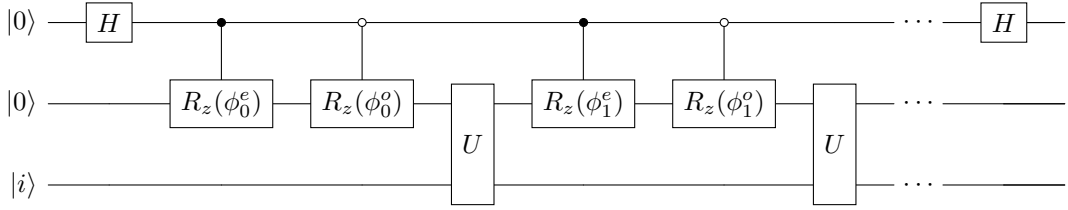
\begin{figure}[t]
\centering
\captionsetup{width=0.7\textwidth}
\begin{quantikz}[thin lines]
    \lstick{$\ket{0}$} & \gate{H} & \ctrl{1} & \ctrl[open]{1} &  & \ctrl{1} & \ctrl[open]{1} & & \ \cdots\ & \gate{H} & \qw \\
    \lstick{$\ket{0}$} & & \gate{R_z(\phi_0^{e})} & \gate{R_z(\phi_0^{o})} & \gate[2]{U} & \gate{R_z(\phi_1^{e})} & \gate{R_z(\phi_1^{o})} & \gate[2]{U} &\ \cdots\ & & \qw \\
    \lstick{$\ket{i}$} & & &  &  & & & & \ \cdots\ & & \qw 
\end{quantikz}
\caption{Implementation of $U_\Phi$. When the overall degree is even, the last unitary should be controlled by the first ancilla qubit.}\label{fig2}
\end{figure}

Now we prepare the following $(1,1,0)$-block encoding of $A=\sum_i\cos(x^{(i)})\ket{i}\bra{i}$, which can be implemented by the application of $n$ controlled-$X$ rotations as in Fig. \ref{fig1}:
\begin{align}
    U=\sum_i
    \begin{pmatrix}
    \cos\left(x^{(i)}\right)&i\sin\left(x^{(i)}\right)\\
    i\sin\left(x^{(i)}\right)&\cos\left(x^{(i)}\right)\\
    \end{pmatrix}\otimes\ket{i}\bra{i}.\label{BE}
\end{align}
Using this block-encoding, we could implement the circuit in Fig. \ref{fig2}. Because the operations on the second ancilla qubit are equivalent to QSP, this circuit represents
\begin{align}
    U_\Phi=\sum_i\ket{+}\bra{+}\otimes
    \begin{pmatrix}
    \frac{\tilde{f}^e(x^{(i)})}{|\tilde{f}|_{\mathrm{max}}}&\cdot\\
    \cdot&\cdot\\
    \end{pmatrix}\otimes\ket{i}\bra{i}+\sum_i\ket{-}\bra{-}\otimes
    \begin{pmatrix}
    \frac{\tilde{f}^o(x^{(i)})}{|\tilde{f}|_{\mathrm{max}}}&\cdot\\
    \cdot&\cdot\\
    \end{pmatrix}\otimes\ket{i}\bra{i}.
\end{align}
To implement this circuit, we use $U$ and $U^\dag$ $d/2$ times each.
Applying $U_\Phi$ to an initial state $\ket{00}\ket{+}^{\otimes n}$, we get
\begin{align}
    \ket{\Psi_{\tilde{f}}}&=\sum_i\frac{\tilde{f}(x^{(i)})}{2\sqrt{N}|\tilde{f}|_{\mathrm{max}}}\ket{00}\ket{i}+\ket{\perp}\\
    &=\frac{\mathcal{F}_{\tilde{f}}}{2}\ket{00}\ket{\psi_{\tilde{f}}}+\ket{\perp},
\end{align}
where $\mathcal{F}_{\tilde{f}}=\mathcal{N}_{\tilde{f}}/(\sqrt{N}|\tilde{f}|_{\mathrm{max}})$.
Then we can apply exact amplitude amplification $O(1/\mathcal{F}_{\tilde{f}})$ times to obtain the target state $\ket{\psi_{\tilde{f}}}$.\\

\begin{theorem}[Single variable state preparation, Ref. \cite{mcardle2022quantum}]
Given a polynomial $\tilde{f}$ of degree $d_\delta$ that approximates $f$ to $L_{\infty^{-}}$error $\delta=\epsilon \cdot \operatorname{Min}\left(\mathcal{F}_f, \mathcal{F}_{\tilde{f}}\right)$, we can prepare a quantum state $\ket{\psi_{\tilde{f}}}$ that is $\epsilon$-close in trace-distance to $\ket{\psi_f}$ using $O\left(nd_\delta/\mathcal{F}_{\tilde{f}}\right)$ gates.
\label{SingleState}
\end{theorem}

\bibliography{cite}
\bibliographystyle{junsrt}
\end{document}